\documentclass[12pt]{article} 
\RequirePackage[colorlinks,citecolor=blue,urlcolor=blue]{hyperref}
\usepackage[utf8]{inputenc} 

\usepackage{geometry} 
\usepackage{graphicx} 

\usepackage{tikz, tikz-qtree}
\usepackage{dsfont}
\usetikzlibrary{decorations.pathreplacing}
\usepackage[edges]{forest}
\usepackage{booktabs} 
\usepackage{array} 
\usepackage{paralist} 
\usepackage{verbatim} 
\usepackage{subfig}
\usepackage{floatrow}
\usepackage{amsbsy}
\usepackage{amsmath}
\usepackage{amssymb}
\usepackage{amsfonts}
\usepackage{amscd}
\usepackage{amsthm}
\usepackage{mathabx}
\usepackage{dutchcal}
\usepackage{float}
\usepackage[toc,page]{appendix}
\usepackage{natbib}
\usepackage{mathrsfs}
\usepackage{lineno}
\usepackage{todonotes}
\usepackage{setspace}
\usepackage{multicol}
\usepackage{multirow}
\usepackage{algorithm}
\usepackage{algorithmicx}
\usepackage{algpseudocode}
\usepackage{fancyhdr} 
\usepackage{sectsty}
\allsectionsfont{\sffamily\mdseries\upshape} 

\usepackage[nottoc,notlof,notlot]{tocbibind} 
\usepackage[titles,subfigure]{tocloft} 

\graphicspath{{Figures/}}

\newcommand{\order}[1]{{\cal O}\hspace{-0.2em}\left( #1 \right)}
\newcommand\Tstrut{\rule{0pt}{2.6ex}}         
\renewcommand{\Pr}[0]{\mathbb{P}}
\newcommand{\phylogeny}[0]{\mathcal{F}}

\newcommand{\nTips}[0]{N}

\newcommand{\momentum}[0]{\mathbf{p}}

\newcommand{\parent}[1]{\mbox{\small pa}\hspace{-0.1em}\left( #1 \right)} 

\newcommand{\gradient}[1]{\frac{\partial}{\partial #1}}

\newcommand{\lnP}[1]{\ensuremath{\log \Pr ({#1})}}
\newcommand{\bl}[1]{{b}_{#1}} 
\newcommand{\br}[1]{\mathcal{r}_{#1}} 
\newcommand{\nodeT}[1]{t_{#1}} 
\newcommand{\mass}{\mathbf{M}}
\newcommand{\transpose}{^{\prime}}
\newcommand{\epochLength}[1]{L_{#1}}
\newcommand{\nodeHeight}[1]{t_{#1}}
\newcommand{\epochMap}[1]{\mathcal{E}{(#1)}}
\newcommand{\epochNode}[2]{{#1}_{#2}}
\newcommand{\epochNodeNum}[1]{m_{#1}}
\newcommand{\ratio}[1]{r_{#1}}
\newcommand{\ratioProd}[1]{S_{#1}}
\newcommand{\Jacobian}[0]{\mathbf{J}}

\begin{document}
\begin{flushright}
Version dated: \today
\end{flushright}

\bigskip
\medskip

\begin{center}
	\begin{Large}
		{\bf Scalable Bayesian divergence time estimation with ratio transformations}
	\end{Large}
\end{center}


\begin{center}
	{Xiang Ji$^{1}$, Alexander A. Fisher$^{2}$, Shuo Su$^{3}$, Jeffrey L. Thorne$^{4, 5, 6}$, Barney Potter$^{7}$, Philippe Lemey$^{7}$, Guy Baele$^{7}$, Marc A. Suchard$^{\ast, 8, 9, 10}$\\[2em]
	\small $^{1}$Department of Mathematics, School of Science \& Engineering, Tulane University, \\ New Orleans, LA, USA \\
	\small $^{2}$Department of Statistical Science, Duke University, Durham, NC, USA \\
	\small $^{3}$MOE International Joint Collaborative Research Laboratory for Animal Health \& Food Safety, Jiangsu Engineering Laboratory of Animal Immunology,
	Institute of Immunology, College of Veterinary Medicine, Nanjing Agricultural University,
	Nanjing, Jiangsu, China\\
	\small $^{4}$Bioinformatics Research Center,
	\small $^{5}$Department of Statistics,
	\small $^{6}$Department of Biological Sciences,
	North Carolina State University, Raleigh, NC, USA\\
	\small $^{7}$Department of Microbiology, Immunology and Transplantation, Rega Institute,
	KU Leuven, Leuven, Belgium\\
	\small $^{8}$Department of Biomathematics,
	\small $^{9}$Department of Human Genetics, David Geffen School of Medicine,
	\small $^{10}$Department of Biostatistics, Fielding School of Public Health,
	University of California Los Angeles, Los Angeles, CA, USA \\
	\small $^{\ast}$Correspondence: msuchard@ucla.edu
}\end{center}

\date{}

\clearpage

\begin{abstract}
Divergence time estimation is crucial to provide temporal signals for dating biologically important events, from species divergence to viral transmissions in space and time.
With the advent of high-throughput sequencing, recent Bayesian phylogenetic studies have analyzed hundreds to thousands of sequences.
Such large-scale analyses challenge divergence time reconstruction by requiring inference on highly-correlated internal node heights that often become computationally infeasible.
To overcome this limitation, we explore a ratio transformation that maps the original $\nTips - 1$ internal node heights into a space of one height parameter and $\nTips - 2$ ratio parameters.
To make analyses scalable, we develop a collection of linear-time algorithms to compute the gradient and Jacobian-associated terms of the log-likelihood with respect to these ratios.
We then apply Hamiltonian Monte Carlo sampling with the ratio transform in a Bayesian framework to learn the divergence times in four pathogenic virus phylogenies: West Nile virus, rabies virus, Lassa virus and Ebola virus.
Our method both resolves a mixing issue in the West Nile virus example and improves inference efficiency by at least 5-fold for the Lassa and rabies virus examples.
Our method also makes it now computationally feasible to incorporate mixed-effects molecular clock models for the Ebola virus example, confirms the findings from the original study and reveals clearer multimodal distributions of the divergence times of some clades of interest.
\end{abstract}


\clearpage

\section{Introduction}
Since \cite{zuckerkandl1962molecular} proposed the first molecular clock model, the development of more reliable divergence time estimation
techniques has thrived.
Because evolutionary rate and time are confounded
in stochastic models for molecular sequence data, divergence time inference can
be improved either via advances in treatment of rates or treatment of times.
However, the
majority of the effort has centered upon improving the model aspects that describe how evolutionary rates change across the tree while
the other confounding component --- the evolutionary times --- has received
less attention.

This imbalance is partly due to the constraints on the node heights imposed by the tree structure.
Assuming a rooted tree with the root node on the top and tip nodes at the bottom, an internal node must be higher than its descendant nodes but lower than its parent node.
These constraints pose great challenge  for inferring internal node heights jointly, such that one typically samples or optimizes the height of one node at a time.

Despite this inference difficulty, divergence time estimation is crucial to provide temporal signals for dating biologically important events, from species divergence to viral transmissions in space and time \citep{erwin2011cambrian, meredith2011impacts, dux2020measles, lemey2020accommodating}.
Repeated breakthroughs in sequencing technologies have led to molecular data
accumulating at an ever-increasing pace.
The result is often data sets that contain so many sequences that desired
divergence time analyses become computationally
infeasible.  When faced with such obstacles, investigators resort
to analyzing only a small proportion of the
available data and/or sacrificing statistical
rigor and biological plausibility by adopting
procedures and models that are flawed
but computationally convenient.  There is therefore
substantial value in reducing the amount of
computation necessary for statistically sound
divergence time inference.


In \cite{kishino2001performance}, the authors transform the internal node heights of a phylogeny with contemporaneous data sampled at the same time into a collection of ratios that sum to $1$.
With a Dirichlet prior distribution, \citeauthor{kishino2001performance} were then able to jointly sample all proportions at one time.
Inspired by their pioneering work, we explore a more general ratio transformation, similar to that used in \cite{fourment2019evaluating}, for the internal node heights that one can apply to both serially sampled or contemporaneous data.
The ratio transformation serves as a reparameterization that works with any existing phylogenetic models without need for any specific prior.
In fact, the proposed ratio transformation keeps the topology-imposed constraints by its construction with the ratios being independent such that they
are easy to sample from or optimize on.


We show that one can calculate the transformation and the determinant of the Jacobian matrix of the transformation in linear-time with respect to (w.r.t) the number of tips ($\nTips$).
With the determinant of the Jacobian matrix, one can set up the phylogenetic model w.r.t.~the untransformed node heights, but sample from the transformed ratio space.
To make use of an advanced linear-time gradient of the log-likelihood algorithm \citep{ji2020gradients}, we show that one can transform the gradient w.r.t.~the untransformed node heights to the gradient w.r.t.~the transformed ratio space with $\order{\nTips}$ calculations.
The linear-time gradient transformation enables the application of gradient-based Monte Carlo samplers such as the Hamiltonian Monte Carlo (HMC) method \citep{neal2011mcmc} in the Bayesian framework.
HMC shows great potential for improving computational efficiency in many phylogenetic applications \citep{dinh2017probabilistic, ji2020gradients, baele2020hamiltonian}.

We apply the ratio transformation to simultaneously learn the branch-specific evolutionary rates and the internal node heights of four viral examples.
Our method significantly improves inference efficiency with a 5- to 8-fold computational performance increase for our Lassa and rabies virus examples.
More interestingly, our West Nile virus example shows that our sampler
better approximates
the posterior density than do classic univariate samplers that suffer from
Markov chain Monte Carlo (MCMC) mixing issues.
For an Ebola virus example, we show that our method makes it computationally
feasible to employ a relaxed clock model to account for both clade- and branch-specific effects that reveal clearer multi-modal distribution
of times for clades of interest.

\section{New Approach} \label{sec:algorithm}
In this section, we define necessary notation for deriving the ratio transformation.
We then illustrate the transformation from the node heights into the ratio space and its reverse transform.
We show that the Jacobian matrix of the transform is a triangular matrix such that its determinant calculation only involves the diagonal elements.
We derive linear-time algorithms that transform the gradient of the sequence data log-likelihood w.r.t.~the node heights into the gradient w.r.t.~ the ratio space with complete post- and pre-order traversals.
We finish the section with another linear-time algorithm that calculates the gradient of the log-Jacobian term w.r.t.~the ratio space.

\subsection{Notation}
Consider a
rooted phylogeny $\phylogeny$ with $\nTips$ tips and $\nTips-1$ internal nodes.
Assume that the root node is on the top and the tip nodes are at the bottom of $\phylogeny$.
We denote the tip nodes with numbers $1, 2, ..., \nTips$ and the internal nodes with numbers $\nTips+1, \nTips+2, ..., 2\nTips-1$ where the root node is $2\nTips-1$.
We denote $\parent{i}$ as the parent node of node $i$.
Any branch on the tree is denoted by the number of the child node that ends it (i.e.~branch $i$ connects node $\parent{i}$ to $i$).
We denote $\nodeT{i}$ as the height (i.e., time) of node $i$.
When $i$ is a tip node (i.e., $i \in \{1, 2, \dots, \nTips\}$), its height is the sampling time.
In divergence time estimation, one is interested in estimating the heights of internal nodes.


Without loss of generality, we derive the ratio transform where the tip
nodes can be associated with
serially sampled data
and where the
transformation with contemporaneous data is then a special case where
all tip node times are identical.
We first define epochs such that any internal node belongs to one and only one epoch.
We then define a
ratio parameter ascribed to each of the internal nodes except for the root.

\subsection{Epoch construction and the ratio transformation} 
\begin{figure}[h!]
  \begin{center}
  \includegraphics[width=0.8\textwidth]{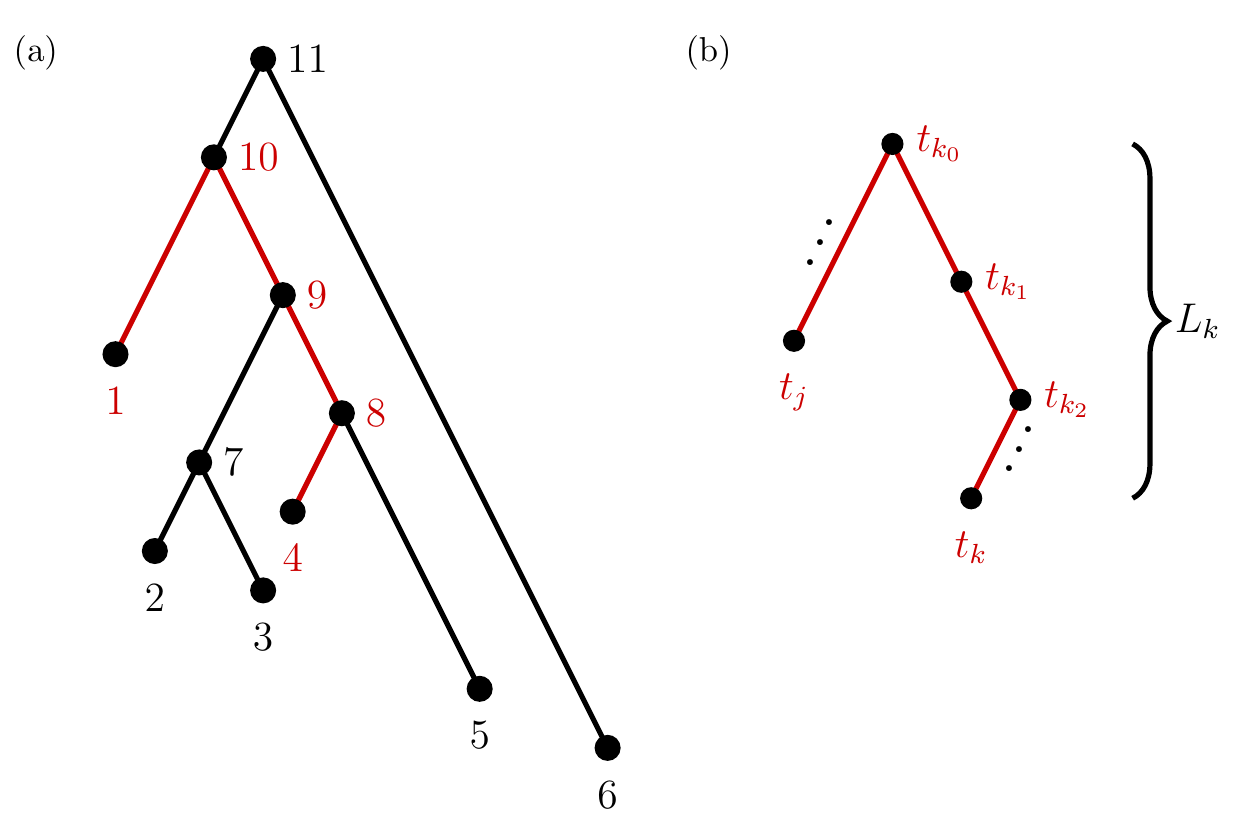}
	\caption{Epoch construction on a $6$-taxa tree.  (a) Example tree with serially sampled data.  (b) One epoch example. For the example tree in (a) with anchor tip $4$, $k=4$, $j=1$ and $k_0=10$.
		For anchor tip $2$, $k=2$, $j=4$ and $k_0=9$.  For anchor tip $1$, $k=1$ is the starting epoch that contains the root node.}
	\label{fig:Epoch}
	\end{center}
\end{figure}

For an internal node, we refer to its earliest (i.e.~highest) descendant tip node as its \textit{anchor node}.
Therefore, the anchor node of an internal node is its closest descendant tip node.
To make the anchor nodes consistent and unique, we assign an arbitrary ordering among tip nodes to distinguish those with the same sampling times.
For example, we pick the tip node with the smallest node number as the anchor node from all closest tip nodes sampled at the same time.
We group all internal nodes with the same anchor node into an epoch.
We refer to an epoch by the number of its anchor node.
Except for the epoch to which the root node belongs, an epoch is
constructed to have
a chain structure from its anchor node up to the highest node in the epoch (see Figure~\ref{fig:Epoch}).
We refer to the parent node of the highest node in an epoch as its \textit{connecting node} such that the connecting node of an epoch belongs to another epoch.
We treat the root node as the connecting node for epochs of its immediate descendant nodes.

Let $\nodeHeight{i}$ denote the node height of node $i$ and $\epochMap{i}$ be the epoch to which node $i$ belongs.
We refer to the epoch to which the root node belongs as the starting epoch and assign it as $\epochMap{2N - 1}$.
We abuse notation by referring to the $j^{\mbox{\tiny th}}$ node of epoch $k$ as $\epochNode{k}{j}$.
For epoch $k$ that contains $\epochNodeNum{k}$ internal nodes with strictly positive branch lengths, we have $\{\nodeHeight{k_1}, \nodeHeight{k_2}, ..., \nodeHeight{k_{m_k}}:  \nodeHeight{k_1} > \nodeHeight{k_2} > ... > \nodeHeight{k_{m_k}}  > \nodeHeight{k} \}$.
We refer to the connecting node of an epoch as the $0^{\mbox{\tiny th}}$ node of an epoch, i.e.~$k_0 = \parent{k_1}$.
We define $\epochLength{k}=\nodeHeight{k_0} - \nodeHeight{k}$ as the length of epoch $k$.
For the $i^{\mbox{\tiny th}}$ internal node $k_i$ from epoch $k$ (i.e.~$i > 0$), we define its ratio parameter $\ratio{k_i}$ as
\begin{equation}
	\label{eq:ratio-transform}
	\ratio{k_i} = \frac{\nodeHeight{k_i} - \nodeHeight{k}}{\nodeHeight{k_{i-1}} - \nodeHeight{k}},
\end{equation}
where $t_k$ is the height of the anchor node of epoch $k$ and $k_{i - 1} = \parent{k_i}$.
Therefore, one can write the time of an internal node as a function of the ratios and the epoch lengths as
\begin{equation}
	\label{eq:inverse-ratio}
	\nodeHeight{k_i} = \epochLength{k} \prod\limits_{n=1}^{i} {\ratio{k_n}} + \nodeHeight{k}.
\end{equation}
To ease notation, let $\ratioProd{k_i} = \prod\limits_{n=1}^{i} {\ratio{k_n}}$ be the product of ratios for internal node $k_i$ of epoch $k$.
Equation~\ref{eq:inverse-ratio} simplifies to
\begin{equation}
	\label{eq:time_short}
	\nodeHeight{k_i} = \epochLength{k} \ratioProd{k_i} + \nodeHeight{k}.
\end{equation}
%
Interestingly, there is only one degree of freedom for all epoch lengths because
\begin{equation}
	\label{eq:epochLength}
	\begin{aligned}
		\nodeHeight{k_0}
		&= \nodeHeight{k} + \epochLength{k} \\
		&= \nodeHeight{\epochMap{k_0}} + \epochLength{\epochMap{k_0}} \ratioProd{k_0},
	\end{aligned}
\end{equation}
such that
the length of epoch $k$ is determined by the length of the epoch of its connecting node and the two associated anchor node times.
We arrive at the following recursive relationship for epoch lengths
\begin{equation}
	\label{eq:epochLengthRecursion}
	\begin{aligned}
		\epochLength{k} &= \nodeHeight{\epochMap{k_0}} - \nodeHeight{k} + \epochLength{\epochMap{k_0}} \ratioProd{k_0}\\
	\end{aligned} .
\end{equation}
Therefore, there is effectively only one degree of freedom for the scale of time with all ratios denoting the relative height an internal node has using its parent node and the anchor node as reference.
There are many choices for modeling this single dimension for time scale, e.g., one may arbitrarily choose one of the epoch lengths.
We pick the starting epoch length as the free parameter $\epochLength{\epochMap{2\nTips - 1}} = \nodeHeight{2\nTips - 1} - \nodeHeight{\epochMap{2\nTips - 1}}$ which we refer to as the \textit{height parameter} because it represents the height difference from the root node to its closest tip node (all tip nodes are descendants of the root).
We refer to the space of the height and $\nTips - 2$ ratio parameters as the \textit{ratio space}.
We refer to the space of all untransformed node heights as the \textit{height space}.
We refer to the transformation from
the height space into the ratio space as the \textit{ratio transform}.

Algorithm~\ref{alg:ratio_transform} illustrates the ratio transform through a single post-order traversal that visits every node on the tree in a descendant-first manner.
Likewise, one can perform the inverse ratio transform to get node heights from the ratios by reversing Equation~\ref{eq:ratio-transform} through a pre-order traversal.
\begin{algorithm}[h!]
	\caption{Ratio transform through a single post-order traversal}
	\label{alg:ratio_transform}
	\begin{algorithmic}
		\For{node $i$ in a post-order traversal}
		\If{$i$ is a tip node}
		\State{Set the anchor tip of epoch $i$ as node $i$.}
		\Else
		\State{Set the anchor tip of $i$ the same as the anchor tip of its highest daughter node.}
		\State{Calculate $\ratio{i}$ according to Equation~\ref{eq:ratio-transform}.}
		\EndIf
		\EndFor
	\end{algorithmic}
\end{algorithm}

\subsection{Gradient and Jacobian}

Many modern inference machineries benefit from gradient information to find descending directions of the likelihood surface or to efficiently integrate dynamics along the surface for generating Monte Carlo proposals (e.g.~\cite{ji2020gradients} contains gradient applications in non-linear optimization and Bayesian posterior sampling).
When transforming probability densities from their original space into another (e.g.~the ratio space in this case), one needs the determinant of the Jacobian matrix to correctly ``weight'' the transformed density (see~Theorem 2.1.5 from \cite{casella2001statistical}).
In this section, we derive algorithms for transforming the ``unweighted'' likelihood into the ratio space together with the associated quantities from the log-determinant of the Jacobian matrix to correctly set the ``weight''.

In \cite{ji2020gradients}, we introduced a linear-time algorithm for calculating the gradient of the log-likelihood w.r.t.~the branch length $\bl{i} = \br{i} (\nodeT{i} - \nodeT{\parent{i}})$ that is the product of the evolutionary rate $\br{i}$ and the time duration $\nodeT{i} - \nodeT{\parent{i}}$ of branch $i$.
To calculate the gradient w.r.t.~node heights, one starts with the gradient w.r.t.~branch lengths and finishes via the chain rule.
More specifically, for node $i$ with its two immediate descendant nodes $j$ and $k$, the derivative of the log-likelihood, $\lnP{\mathbf{Y}}$, w.r.t.~$\nodeT{i}$ is:
\begin{equation}\label{eq:NodeHeightGradient}
	\begin{aligned}
		\gradient{\nodeHeight{i}}{\lnP{\mathbf{Y}}}
		=
		\left\{ {
			\begin{array}{*{20}{l}}
				{\frac{\partial \lnP{\mathbf{Y}}}{\partial \bl{i}}\frac{\partial \bl{i}}{\partial \nodeT{i}} + \frac{\partial \lnP{\mathbf{Y}}}{\partial \bl{j}}\frac{\partial \bl{j}}{\partial \nodeT{i}} + \frac{\partial \lnP{\mathbf{Y}}}{\partial \bl{k}}\frac{\partial \bl{k}}{\partial \nodeT{i}}}, &{\mbox{$i \ne 2\nTips - 1$}}\\
				{\frac{\partial \lnP{\mathbf{Y}}}{\partial \bl{j}}\frac{\partial \bl{j}}{\partial \nodeT{i}} + \frac{\partial \lnP{\mathbf{Y}}}{\partial \bl{k}}\frac{\partial \bl{k}}{\partial \nodeT{i}}}, &{\mbox{$i = 2\nTips - 1$}}.
		\end{array}} \right. \\
	\end{aligned}
\end{equation}
It is important to recall that a ratio parameter is only explicit to the node it assigns to and all its descendant nodes by Equation~\ref{eq:inverse-ratio}.
Therefore, we only need the partial derivatives $\frac{\partial \nodeHeight{k}}{\partial \ratio{i}}$ from node $i$ and all its descendant nodes to finish the chain rule
\begin{equation}\label{eq:gradientChainRule}
	\gradient{\ratio{i}}{\lnP{\mathbf{Y}}} = \sum\limits_{k}{\left[ \gradient{\nodeHeight{k}}{\lnP{\mathbf{Y}}}\frac{\partial \nodeHeight{k}}{\partial \ratio{i}} \right] }.
\end{equation}
To derive the partial derivative $\frac{\partial \nodeHeight{k}}{\partial \ratio{i}}$ for any two nodes $i$ and $k$ such that node $k$ is a
descendant of node $i$, we separate the node pairs into two cases.
The first case considers node $i$ and node $k$ in the same epoch (including the pair where $i = k$, e.g.~Equation~\ref{eq:time_short}), such that
\begin{equation}\label{eq:gradientSameEpoch}
	\begin{aligned}
		\frac{\partial \nodeHeight{k}}{\partial \ratio{i}}
		=&  \epochLength{\epochMap{k}} \frac{\partial \ratioProd{k} }{\partial \ratio{i}}\\
		=&  \frac{ \nodeHeight{k} - \nodeHeight{\epochMap{k}} }{\ratio{i}}. \\
	\end{aligned}
\end{equation}
%
For the other case where node $i$ and node $k$ belong to different epochs, we start with revealing the relationship between the partial derivatives of node $k$'s height $\nodeHeight{k}$ and its connecting node $\epochMap{k}_0$'s height $\nodeHeight{\epochMap{k}_0}$ w.r.t.~the same ratio $\ratio{i}$ (e.g.~plug Equation~\ref{eq:epochLengthRecursion} in Equation~\ref{eq:time_short}), such that
\begin{equation}
	\label{eq:derivativeInductiveRule}
	\begin{aligned}
		\frac{\partial \nodeHeight{k}}{\partial \ratio{i}}
		=& \ratioProd{k} \frac{\partial \left( \nodeHeight{\epochMap{k}_0} - \nodeHeight{\epochMap{k}} + \epochLength{\epochMap{i}}\ratioProd{\epochMap{k}_0} \right) }{\partial \ratio{i}} \\
		=& \ratioProd{k} \frac{\partial \nodeHeight{\epochMap{k}_0}}{\partial \ratio{i}}.\\
	\end{aligned}
\end{equation}
Equation~\ref{eq:derivativeInductiveRule} shows that one obtains the partial derivative of a node height $\nodeHeight{k}$ w.r.t.~ratio $\ratio{i}$ by
multiplying the related ratio product (i.e.~$\ratioProd{k}$)
and the partial derivative of the node height $\nodeHeight{\epochMap{k}_0}$ w.r.t.~ratio $\ratio{i}$ (i.e.~$\frac{\partial \nodeHeight{\epochMap{k}_0}}{\partial \ratio{i}}$).
Combining Equations~\ref{eq:gradientSameEpoch} and \ref{eq:derivativeInductiveRule}, we inductively derive a general expression for the derivatives where node $i$ and node $k$ do not belong to the same epoch.
We arrive at this derivation through the existence of a series of connecting nodes (when traveling from node $k$ to node $i$) starting from epoch $\epochMap{k}$ that the last connecting node belongs to the same epoch as node $i$, i.e.~$\epochMap{\epochMap{\dots \epochMap{k}_0}_0}=\epochMap{i}$. 
%
%
The general expression for the derivative becomes
\begin{equation}
	\label{eq:gradientDifferentEpoch}
	\begin{aligned}
		\frac{\partial \nodeHeight{k}}{\partial \ratio{i}}
		=&  \ratioProd{k}  \ratioProd{\epochMap{k}_0} \dots \ratioProd{\epochMap{\dots \epochMap{k}_0}_0} \frac{\partial \nodeHeight{\epochMap{\dots \epochMap{k}_0}_0}}{\partial \ratio{i}}.\\
	\end{aligned}
\end{equation}

By naively plugging Equation~\ref{eq:gradientSameEpoch} and Equation~\ref{eq:gradientDifferentEpoch} into Equation~\ref{eq:gradientChainRule}, we obtain the gradient w.r.t.~the ratio space.
However, this operation amounts to $\order{\nTips^2}$ computations for transforming the gradient. 
To overcome this computational burden, we develop a linear-time $\order{\nTips}$ algorithm for transforming the gradient.

\subsubsection{Post-order traversal}
Consider three internal nodes $i$, $j$ and $k$ such that node $k$ is the parent node of node $i$ and node $j$.
The linear-time algorithm for transforming the gradient w.r.t.~ratio parameters
builds on two properties of the ratio transformation.
The first property is that any descendant node of node $k$ except node $i$ or node $j$ is a descendant node of either node $i$ or node $j$ (for bifurcating trees).
The other property is that node $k$ belongs to the same epoch as either node $i$ or node $j$.
As is
common in dynamic programming algorithms, we want to derive
the relationship of $ \frac{\partial \nodeHeight{l}}{\partial \ratio{k}}$ with $ \frac{\partial \nodeHeight{l}}{\partial \ratio{i}}$ and $ \frac{\partial \nodeHeight{l}}{\partial \ratio{j}}$ where node $l$ is
descendant of node $k$ to reuse quantities cached from evaluating Equation~\ref{eq:gradientChainRule} on descendant nodes.
More specifically, we want to reuse the summations already
determined for $\gradient{\ratio{i}}{\lnP{\mathbf{Y}}}$ and $\gradient{\ratio{j}}{\lnP{\mathbf{Y}}}$ when calculating $\gradient{\ratio{k}}{\lnP{\mathbf{Y}}}$ as in Equation~\ref{eq:derivativeInductiveRule}.

Without loss of generality, we assume node $k$ belongs to the same epoch as node $i$.
The following relationships between derivatives w.r.t.~the three ratio parameters $\ratio{i}$, $\ratio{j}$ and $\ratio{k}$ enable the linear-time algorithm through a single post-order traversal to update the gradient from the node height space into the ratio space.
From Equation~\ref{eq:gradientSameEpoch} and Equation~\ref{eq:gradientDifferentEpoch}, when node $l$ is a descendant of node $i$ (including $i = l$) such that node $k$ and node $l$ are in the same epoch,
\begin{equation}\label{eq:kiPartialRelation}
	\begin{aligned}
		\frac{\partial \nodeHeight{l}}{\partial \ratio{k}}
		&= \frac{\partial \nodeHeight{l}}{\partial \ratio{i}} \frac{\ratio{i}}{\ratio{k}}.\\
	\end{aligned}
\end{equation}
%
When node $l$ is descendant of node $j$ (including $j=l$) such that node $k$ is the connecting node to the epoch $\epochMap{j}$ where node $j$ is the first node,
\begin{equation}\label{eq:kjPartialRelation}
	\begin{aligned}
		\frac{\partial \nodeHeight{l}}{\partial \ratio{k}}
		&= \frac{\partial \nodeHeight{l}}{\partial \ratio{j}} \frac{\ratio{j}}{\epochLength{\epochMap{j}}}
		\frac{\partial \nodeHeight{k}}{\partial \ratio{k}}.\\
	\end{aligned}
\end{equation}
Note that we model the ratio parameters as independent of each other, i.e.~$\frac{\partial \ratio{k}}{\partial \ratio{i}} = \frac{\partial \ratio{k}}{\partial \ratio{j}} = 0$.
Equation~\ref{eq:kiPartialRelation} and Equation~\ref{eq:kjPartialRelation} come from the special structure of the transform that the node height of an internal node is a product of a series of ratio parameters with one single height parameter.
Algorithm~\ref{alg:post_order_gradient_update} illustrates updating the gradient w.r.t.~all ratio parameters (except for the height parameter) where one reuses the derivatives of the log-likelihood w.r.t.~two immediate descendant nodes (i.e.~nodes $i$ and $j$) to calculate the derivative of the log-likelihood w.r.t.~the parent node (i.e.~node $k$).
\begin{algorithm}[h!]
	\caption{Transforming the gradient of the log-likelihood w.r.t.~ratio parameters by post-order traversal}
	\label{alg:post_order_gradient_update}
	\begin{algorithmic}
		\For{node $k$ in a post-order traversal}
		\If{$k$ is a tip node}
		\State{Set the gradient of $k$ as 0.}
		\Else
		\State{Let node $i$ and node $j$ be the two immediate descendant nodes of node $k$}
		\State{such that node $i$ and node $k$ belong to the same epoch.}
		\State{Set the gradient of $k$ as }
		\State{\quad $\gradient{\ratio{k}}{\lnP{\mathbf{Y}}} =
			\gradient{\ratio{i}}{\lnP{\mathbf{Y}}}\frac{\ratio{i}}{\ratio{k}}
			+ \gradient{\ratio{j}}{\lnP{\mathbf{Y}}}\frac{\ratio{j}}{\epochLength{\epochMap{j}}}\frac{\partial \nodeHeight{k}}{\partial \ratio{k}}
			+ \gradient{\nodeHeight{k}}{\lnP{\mathbf{Y}}}\frac{\partial \nodeHeight{k}}{\partial \ratio{k}} $.}
		\EndIf
		\EndFor
	\end{algorithmic}
\end{algorithm}

\subsubsection{Pre-order traversal}
We now update the gradient of the log-likelihood w.r.t.~the height parameter which is the only dimension left in the ratio transform.
We use a pre-order traversal to update the gradient in this dimension because the transformation of all internal node heights depends on it.
The update is
\begin{equation}\label{eq:gradientHeightParameter}
	\gradient{\epochLength{\epochMap{2\nTips - 1}}}{\lnP{\mathbf{Y}}} = \sum\limits_{k}{\left[ \gradient{\nodeHeight{k}}{\lnP{\mathbf{Y}}}\frac{\partial \nodeHeight{k}}{\partial \epochLength{\epochMap{2\nTips - 1}}} \right] }.
\end{equation}
Based on Equation~\ref{eq:epochLength}, we calculate all the partial derivatives $\frac{\partial \nodeHeight{k}}{\partial \epochLength{\epochMap{2\nTips - 1}}}$ according to Algorithm~\ref{alg:pre_order_gradient_update} through a single pre-order traversal.
\begin{algorithm}[h!]
	\caption{Transforming gradient of the log-likelihood w.r.t.~the height parameter by pre-order traversal}
	\label{alg:pre_order_gradient_update}
	\begin{algorithmic}
		\For{node $k$ in a pre-order traversal}
		\If{$k$ is the root node}
		\State{Set the derivative of node height $k$ w.r.t.~height parameter as 1  (i.e.~$\frac{\partial \nodeHeight{2N - 1}}{\partial \epochLength{\epochMap{2\nTips - 1}}} = 1$}).
		\Else
		\State{Set the derivative of $k$ as the product of $\ratio{k}$ and the derivative of its parent node}
		\State{w.r.t.~height parameter (i.e.~$\frac{\partial \nodeHeight{k}}{\partial \epochLength{\epochMap{2\nTips - 1}}} = \ratio{k} \frac{\partial \nodeHeight{\mbox{\tiny pa}(k)}}{\partial \epochLength{\epochMap{2\nTips - 1}}}$}).
		\EndIf
		\EndFor
	\end{algorithmic}
\end{algorithm}

\subsubsection{Determinant of the Jacobian matrix}
We now derive the Jacobian matrix associated with the ratio transform whose determinant sets the weight for the transformed density.
One derives the full Jacobian matrix for the ratio transform by applying Equation~\ref{eq:gradientSameEpoch} and Equation~\ref{eq:gradientDifferentEpoch}.
Note the special structure that has $\frac{\partial \nodeHeight{k}}{\partial \ratio{i}} \ne 0$ if and only if $i=k$ or node $k$ is descendant of node $i$, and also
note the independence between the height parameter and the ratio parameters.
By ordering the entries in a descendant node first fashion
that coincides with how nodes are visited in a post-order traversal, the Jacobian matrix becomes triangular (including the height parameter).
Because the determinant of a triangular matrix only involves the diagonal entries, the determinant of the Jacobian matrix $\Jacobian$ becomes
\begin{equation}
	\label{Eq:ratioJacobianDeterminant}
	\begin{aligned}
		|\Jacobian|
		&= \prod\limits_{i} \frac{\partial \nodeHeight{i}}{\partial \ratio{i}} \\
		&= \prod\limits_{i} \left[ \nodeHeight{\parent{i}} - \nodeHeight{\epochMap{i}} \right].\\
	\end{aligned}
\end{equation}

\subsubsection{Gradient of log-determinant of the Jacobian matrix}
We complete this section with a final linear-time algorithm for calculating the gradient of the log-determinant of the Jacobian matrix w.r.t.~the ratio space for applying HMC on this transformed space as described in the next section.
This additional gradient component facilitates using HMC to sample all dimensions jointly in the ratio space.
Similar to the case of updating the gradient of the log-likelihood
from the original space into the ratio space, naively applying Equation~\ref{eq:gradientSameEpoch} and Equation~\ref{eq:gradientDifferentEpoch} results in
an undesired quadratic computational load.
One can benefit from the same properties that lead to Algorithm~\ref{alg:post_order_gradient_update} with a modified two-pass linear-time Algorithm~\ref{alg:post_order_gradient_log_determinant} that calculates all the derivatives of the log-determinant of the Jacobian matrix w.r.t.~the ratio parameters.
\begin{algorithm}[h!]
	\caption{Calculating gradient of the log-determinant of the Jacobian matrix w.r.t.~ratio parameters by post-order traversal}
	\label{alg:post_order_gradient_log_determinant}
	\begin{algorithmic}
		\For{node $k$ in a post-order traversal}
		\If{$k$ is a tip node}
		\State{$\gradient{\ratio{k}}{\log |\Jacobian|} = 0$}
		\Else
		\State{Let node $i$ and node $j$ be the two immediate descendant nodes of node $k$}
		\State{such that node $i$ and node $k$ belong to the same epoch, and compute}
		\State{$\gradient{\ratio{k}}{\log |\Jacobian|} =
			\gradient{\ratio{i}}{\log |\Jacobian|} \frac{\ratio{i}}{\ratio{k}}
			+ \gradient{\ratio{j}}{\log |\Jacobian|}\frac{\ratio{j}}{\epochLength{\epochMap{j}}}\frac{\partial \nodeHeight{k}}{\partial \ratio{k}}
			+ \frac{1}{\nodeHeight{k} - \nodeHeight{\epochMap{k}}} \frac{\partial \nodeHeight{k}}{\partial \ratio{k}} $ .}
		\EndIf
		\EndFor
		\For{every internal node $k$}
		\State{Update $\gradient{\ratio{k}}{\log |\Jacobian|} = \gradient{\ratio{k}}{\log |\Jacobian|} - \frac{1}{\ratio{k}}$ .}
		\EndFor
	\end{algorithmic}
\end{algorithm}
%

\newcommand{\position}{\boldsymbol{\theta}}

\subsection{Hamiltonian Monte Carlo}
HMC is a state-of-the-art MCMC method that generates efficient proposals through Hamiltonian dynamics \citep{neal2011mcmc} for the Metropolis-Hastings algorithm \citep{metropolis53}.
For an arbitrary and unbounded parameter of interest $\position$ with the posterior density $\pi(\position)$, HMC introduces an auxiliary parameter $\momentum$ and samples from the product density $\pi(\position, \momentum) = \pi(\position)\pi(\momentum)$ through:
\begin{equation}
	\label{eq: HMC}
	\begin{aligned}
		\frac{\textrm{d} \momentum}{\textrm{d} t} &=  - \nabla U(\position) = \nabla \log \pi(\position)  \text{ and} \\
		\frac{\textrm{d} \position}{\textrm{d} t} &=  \nabla K(\momentum) = \mass^{-1} \momentum , \\
	\end{aligned}
\end{equation}
where $U(\position)$ is the `potential energy' often set to the negative log-posterior density and $K(\momentum) = \momentum\transpose \mass^{-1} \momentum/2$ is the `kinetic energy' as the auxiliary parameter $\momentum$ typically follows a multivariate normal distribution $\momentum \sim \mathscr{N}\hspace{-0.1em}(\mathbf{0},\mass)$ with a `mass matrix' $\mass$ as the covariance matrix.
HMC has shown great potential in diverse phylogenetic applications \citep{dinh2017probabilistic, ji2020gradients, baele2020hamiltonian}.

Naive application of HMC on the space of internal node heights is highly inefficient because of the irregular constraints on these parameters.
Instead, the ratio space is trivial to extend such that it is unbounded by applying a logit-transform to each ratio independently and a log-transfrom to the single height parameter.
We apply HMC on the (extended) ratio space for efficient sampling of all internal node heights.
Finally, we also apply HMC for jointly sampling the evolutionary rates and times (i.e.~divergence time estimation) and explore the additional efficiency gain this affords.


\paragraph{Preconditioning with adaptive variance}
The geometric structure of the posterior distribution significantly affects the computational efficiency of HMC.
For example, when the scales of the posterior distribution vary among individual parameters, failing to account for such structure may reduce the efficiency of HMC \citep{neal2011mcmc, Stan2017, ji2020gradients}.
We can adapt HMC for such structure by modifying the dynamics in Equation~\ref{eq: HMC} via an appropriately chosen mass matrix $\mass$.
In \cite{ji2020gradients}, we employ a mass matrix informed by the diagonal entries of the Hessian matrix of the log-posterior to account for the variable scales among dimensions.
Unfortunately, one needs the full Hessian matrix in the original height space to transform into the Hessian matrix w.r.t.~the ratio space.
This strategy is too computationally expensive to adopt.

\newcommand{\iterPerUpdate}{k}
To incorporate information from the covariance matrix without excessive computational burden, we seek an alternative adaptive MCMC procedure \citep{haario1999adaptive, Andrieu2008adap_mcmc, roberts2009examples}. Adaptive MCMC has previously found its way into Bayesian phylogenetic inference \citep{baele2017adaptive} and we use this technique here to tune $\mass$ to the covariance matrix estimated from previous samples in the Markov chain.
We further restrict $\mass$ to remain diagonal and hence to scale the ratio dimensions according to their marginal covariance.
This restriction is commonly imposed to regularize the estimate, and a diagonal matrix alone can greatly enhance sampling efficiency of HMC in many situations \citep{Stan2017, pymc16, ji2020gradients}.
We start the HMC sampler with  an identity $\mass$ matrix to collect an initial set of samples (e.g. 200 in our analyses), 
after which we employ the sample covariance to tune $\mass$ adaptively.
Also, we only update the diagonal mass matrix every $10$ HMC iterations so that the cost of computing the adaptive $\mass$ diagonals remains negligible.

\subsection{Emerging viral sequences}

We examine the molecular evolution of West Nile virus (WNV) in North America (1999 - 2007), rabies virus (RABV) in the United States (1982 - 2004), the S segment of Lassa virus (LASV) in West Africa (2008 - 2013) and Ebolavirus (EBOV) in
the
Democratic Republic of Congo, Africa (2018 - 2020) \citep{ Pybus2012, biek2007high, andersen2015clinical, mbala2021ebola}.
In all four virus data sets, phylogenetic analyses have revealed a high variation of the evolutionary rates across branches in the underlying phylogeny.

\paragraph{West Nile virus}

West Nile virus is a mosquito-borne RNA virus that involves multiple species of mosquitoes and birds where birds are the primary host.
WNV first emerged in the Americas in New York in 1999, and quickly spread across the continent, causing an epidemic of human disease accompanied with massive bird deaths.
In total, human infections have resulted in over 48,000 reported cases, 24,000 reported neuroinvasive cases, and over 2,300 deaths \citep{hadfield2019twenty}.
The molecular sequences consist of 104 full genomes, with a total alignment length of 11,029 nucleotides, and were collected from infected human plasma samples from 2003 to 2007 as well as near-complete genomes obtained from GenBank \citep{Pybus2012}.

\paragraph{Rabies virus}
Rabies is an RNA virus that can cause zoonotic disease and is responsible for over $50,000$ human deaths every year.
Besides bats, several terrestrial carnivore species, such as raccoons are important rabies reservoirs.
Before the detection of a raccoon-specific rabies virus variant in 1970s, the limited focus of raccoons as a primary host for rabies was in the southeastern U.S., particularly Florida. 
Over the following decades, an emergence of the virus spread along the mid-Atlantic coast and northeastern U.S.
We analyze the molecular sequences originally described in \cite{biek2007high} that previously served as an example dataset in work on the flexible non-parametric skygrid coalescent model \citep{gill2016understanding}.
The data consist of $47$ sequences sampled from rabid raccoons between 1982 and 2004 that contain the complete rabies nucleoprotein gene (1365 bp) with part of a noncoding region (87 bp) immediately following its 3' end, and a large portion of the glycoprotein gene (1359 bp).

\paragraph{Lassa virus}

Lassa virus is the causative agent of Lassa fever, a hemorrhagic fever endemic to parts of West Africa that is responsible for thousands of deaths and tens-of-thousands of hospitalizations each year \citep{andersen2015clinical}.
LASV infections can lead to Lassa fever, a hemorrhagic fever similar to that from EBOV 
and endemic to parts of West Africa.
Despite the fact that Lassa fever can lead to over 50\% fatality rates among hospitalized patients, an effective vaccine for LASV has yet to be developed and approved.
Unlike EBOV (see next paragraph), 
which passes directly between humans, LASV circulates in a rodent (\emph{Mastomys natalensis}) reservoir and mainly infects humans through contact with rodent excreta.
The LASV genome is comprised of two negative-sense single-stranded RNA segments:
the L segment is $7.3$ kilobase pairs (kb) long, and the S segment is $3.4$ kb long.
In this paper, we use the S segment of the LASV sequence data set of \citet{andersen2015clinical} that consists of $211$ samples obtained at clinics in both Sierra Leone and Nigeria, rodents in the field, laboratory isolates and previously sequenced genomes.

\paragraph{Ebola virus}
The Ebola virus disease (EVD) outbreak in North Kivu province in the Democratic Republic of Congo (DRC) during 2018 - 2020 was the world's second largest Ebola outbreak on record.  It led to 3,481 total cases with 2,299 deaths \citep{DRC2018who}.
One patient who received the recombinant vesicular stomatitis virus-based vaccine was diagnosed with EVD and recovered within 14 days after treatment.  However, 6 months later, the same patient presented again with severe EVD-like illness and EBOV viremia and died \citep{mbala2021ebola}.
The molecular sequences consist of $297$ sequenced isolates that contain $72$
epidemiologically-linked cases to the patient's second infection.

\subsection{Mixed-effects relaxed clock model}
We employ mixed-effects relaxed clock models to learn the evolutionary rates of the four viral datasets as detailed in \cite{bletsa2019}.
More specifically, we use the same random-effects relaxed clock model detailed in \cite{ji2020gradients} for the analysis of WNV, RABV and LASV datasets.
For the EBOV example, we use a mixed-effects relaxed clock model with clade-specific fixed-effects to model clade-specific rate variations among the three branches leading to three clades of interest (relapse clade, MAN14985 clade and KAT21596 clade).
The use of the clade-specific fixed-effects mimics a local clock model that allows us to model and compare possibly different evolutionary rates of persistent infection and overall outbreak, but has previously not been computationally feasible.

\section{Results}

We summarize the computational efficiency improvement with HMC on the ratio space followed by our biological findings on divergence time estimations of the four examples.

\subsection{Computational performance}

We infer the posterior distribution of all internal node heights using two different MCMC transition kernels implemented in BEAST \citep{suchard2018} with likelihood computations off-loaded to the high-performance BEAGLE library \citep{ayres2019beagle} (see Section~\ref{sec:materials_and_methods} for more details). 
The first transition kernel proposes new values for one internal node height at a time from their support.
This represents the current best-practice approach used in BEAST and we will refer to this kernel as ``univariable''.
The other transition kernel utilizes HMC with a diagonal mass matrix informed by adaptive variance on the ratio space that we will refer to as ``HMC''.
As is conventional for Bayesian phylogenetics, we employ a Metropolis-within-Gibbs \citep{tierney1994markov, andrieu2003introduction} inference strategy that cycles between sampling the tree, the evolutionary rates and then other phylogenetic modeling parameters, each from their respective full conditional distributions.

As expected, sampling the topology and the
high-dimensional rate and time (i.e., node height)
parameters is computationally rate-limiting.
Therefore, we explore two scenarios: (1) we sample divergence times only in scenario ``time'';  and (2) we sample evolutionary rate and time jointly in scenario ``rate \& time''.
We compare the efficiency of these transition kernels through their effective sample size (ESS) per unit time for divergence time estimations.
For each analysis, we run the MCMC iterations with each of the transition kernels for roughly the same run time (see Section~\ref{sec:materials_and_methods} for more details of chain lengths from supplementary BEAST XML files).
To maintain identifiability of internal nodes, we constrain the comparisons of
the WNV, RABV and LASV examples to a fixed topology that was randomly selected
from its posterior distribution.  The topologies of the WNV and LASV examples are the same as in \cite{ji2020gradients}.
We relax the topology constraint (i.e.~not fixing the tree topology) for the EBOV example. 
We present the computational efficiency improvement with HMC in the ratio space for sampling node heights.
The application of HMC on the ratio dimensions greatly improves the mixing of the MCMC chain whereas the univariable samplers are problematic for
learning the height of some internal nodes that are
close to the root in the WNV example.
\par
\begin{figure} 
	\begin{center}
		\includegraphics[width=1.0\textwidth]{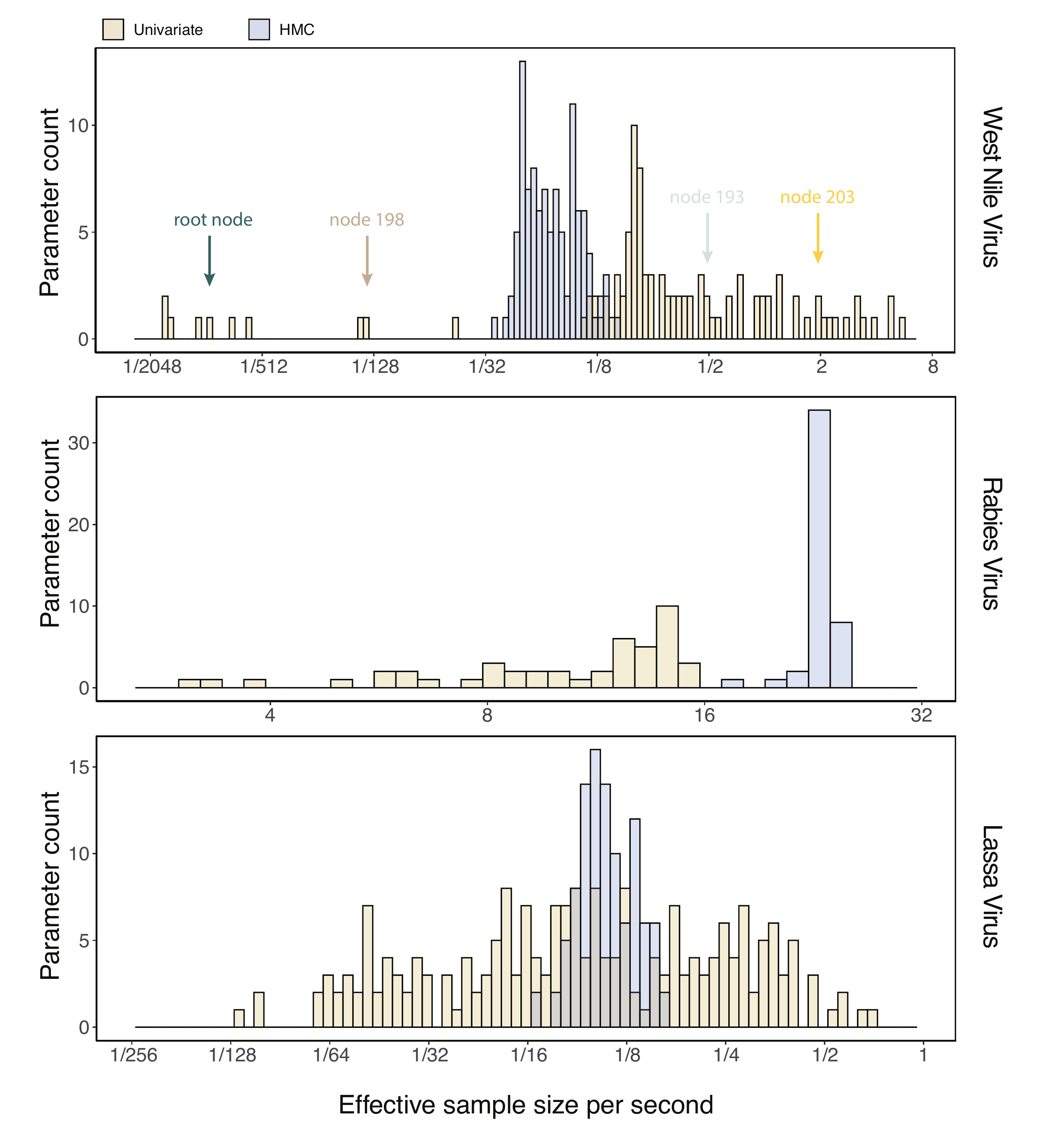}
		\caption{
			Posterior sampling efficiency on all node height parameters for the WNV, RABV, and LASV examples.
			We bin parameters by their ESS/s values.
			The two transition kernels employed in the MCMC are color-coded: a univariable transition kernel and an HMC transition kernel with an adaptive mass matrix.
		}
		\label{fig:ESS}
	\end{center}
\end{figure}

Figure~\ref{fig:ESS} illustrates the posterior sampling efficiency with HMC and univariable samplers in terms of ESS per unit time.
We exclude the WNV example from the efficiency comparison because the poor mixing with univariable samplers leads to an inflated speed-up for HMC.
Table~\ref{tab:ESS} shows the summary statistics of the efficiency gain of the HMC sampler compared with the univariable samplers for the three examples.
The HMC sampler gains at least $5$-fold efficiency improvement in terms of the minimum ESS per unit time in the RABV and LASV examples that
have no difficulties of mixing for the univariable sampler.
\begin{table}
	\caption{
		Computational
		performance of transition kernels for the RABV and LASV examples.
		Computational efficiency measured in terms of effective sample size per second (ESS/s). We compare the performance of our HMC transition kernels operating on the transformed ratio space with a univariable (univariable) transition kernel on the original node height space.
		We report speedup with respect to the minimum and median ESS/s (listed in the columns of ``univariable'' and ``HMC'') across parameters for each example and method.
		We do not report the unreliably high speed-ups for the WNV dataset because of  mixing issues under the ``univariable'' kernel.
	}
	\label{tab:ESS}
	\centering
	\begin{tabular}{lcrrrrrrr}
		\toprule
		\multicolumn{2}{c}{\multirow{2}{*}{ESS/s}}
		& \multicolumn{2}{c}{univariable} & \multicolumn{2}{c}{HMC} & \multicolumn{2}{c}{Speedup}\\
		\cmidrule(lr){3-4} \cmidrule(lr){5-6} \cmidrule(lr){7-8}
		\multicolumn{1}{r}{} & & minimum & median & minimum & median & minimum & median \\
		\hline
		\multirow{2}{*}{RABV} & time &3.187 & 12.154 &17.358 & 23.579 & 5.4$\times$ & 1.9$\times$ \Tstrut\\
		& rate \& time  &0.927 & 4.638 & 6.324 & 8.355 & 6.8$\times$ & 1.8$\times$ \Tstrut\\
		\multirow{2}{*}{LASV} & time &0.008 & 0.090 &0.065 & 0.107 & 7.9$\times$ & 1.2$\times$ \Tstrut\\
		& rate \& time  &0.002 & 0.016 & 0.018 & 0.040 & 8.0$\times$ & 2.4$\times$ \Tstrut\\
		\bottomrule
	\end{tabular}
\end{table}

\subsection{Divergence time estimations}

We summarize divergence time estimation results for each of the viral examples.

\paragraph{West Nile virus}

\begin{figure} 
	\begin{center}
		\includegraphics[width=1.0\textwidth]{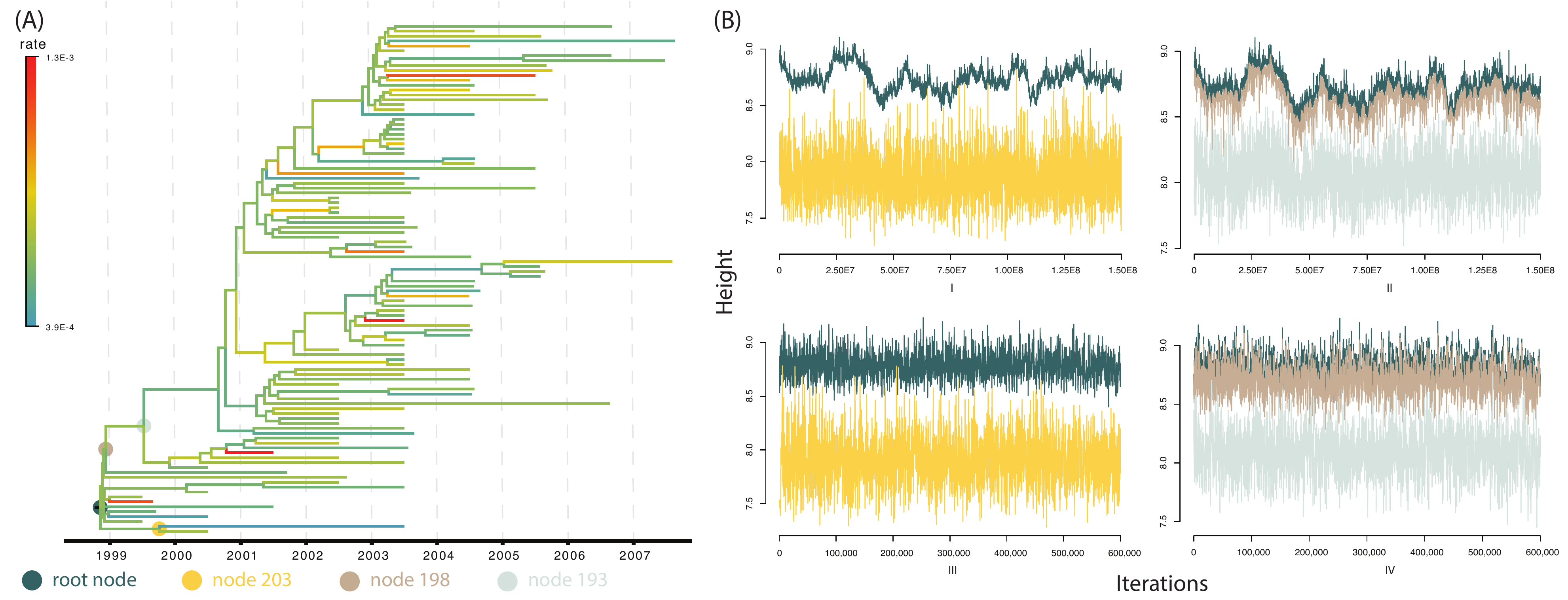}
		\caption{Trace plot of four height parameters indicated on the WNV phylogeny.
			(A) The WNV phylogeny explored in the example.
			Branches on the tree are color-coded by the posterior means of the branch-specific evolutionary rates. 
			Four representative nodes indicated by colored dots on the tree illustrate mixing issues at nodes close to the root when learning the posterior distribution of their heights using the univariable samplers.
			(B) The trace plot of the height parameter of the four nodes indicated in (A) using the same color scheme.  I. and II. are trace plots with the univariable samplers for an MCMC chain of length 1.5 $\times 10^8$ iterations.  III. and IV. are trace plots with the HMC sampler for an MCMC chain of length 600,000.
		}
		\label{fig:TreeTrace}
	\end{center}
\end{figure}

Our analysis estimates the tree-wise (fixed-effect) rate with posterior mean $5.67$ ($95\%$ Bayesian credible interval: $5.05, 6.30$) $\times 10^{-4}$ substitutions per site per year and an estimated variability characterized by the scale parameter of the lognormal distributed branch-specific random-effects with posterior mean $0.34$ $(0.21, 0.47)$.  
These values are similar to previous estimates \citep{Pybus2012, ji2020gradients}.
Figure~\ref{fig:TreeTrace} shows the maximum clade credible evolutionary tree of the WNV example as well as trace plots of several nodes of interest.
Our analysis estimates the date of the epidemic origin to have posterior mean $1998.6$ $(1997.8, 1999.1)$ and this is similar to previous estimates.
Matching previous findings that the American epidemic was likely to originate from the introduction of a single highly pathogenic lineage,
our analysis infers the NY99 lineage to be basal to all other genomes.

Of important note, the MCMC chain suffers from mixing for some height dimensions close to the root (including the root) under the ``univariable'' kernel as illustrated by the traceplot in Figure~\ref{fig:TreeTrace} (B) I and II.
The mixing issue propagates from the root node to a few of its descendant nodes (e.g.~node $198$) 
that plagues over these dimensions because univariable samplers propose a new value for an internal node's height from the interval set by the height of its parent and closest descendant node.
Such a tree-like boundary structure requires multiple height changes on an internal node and the nodes setting its boundaries in the same direction before a ``big'' move is possible that often fails by one of these dimensions moving at the opposite direction.
%

\begin{figure}
	\begin{center}
		\includegraphics[width=1.0\textwidth]{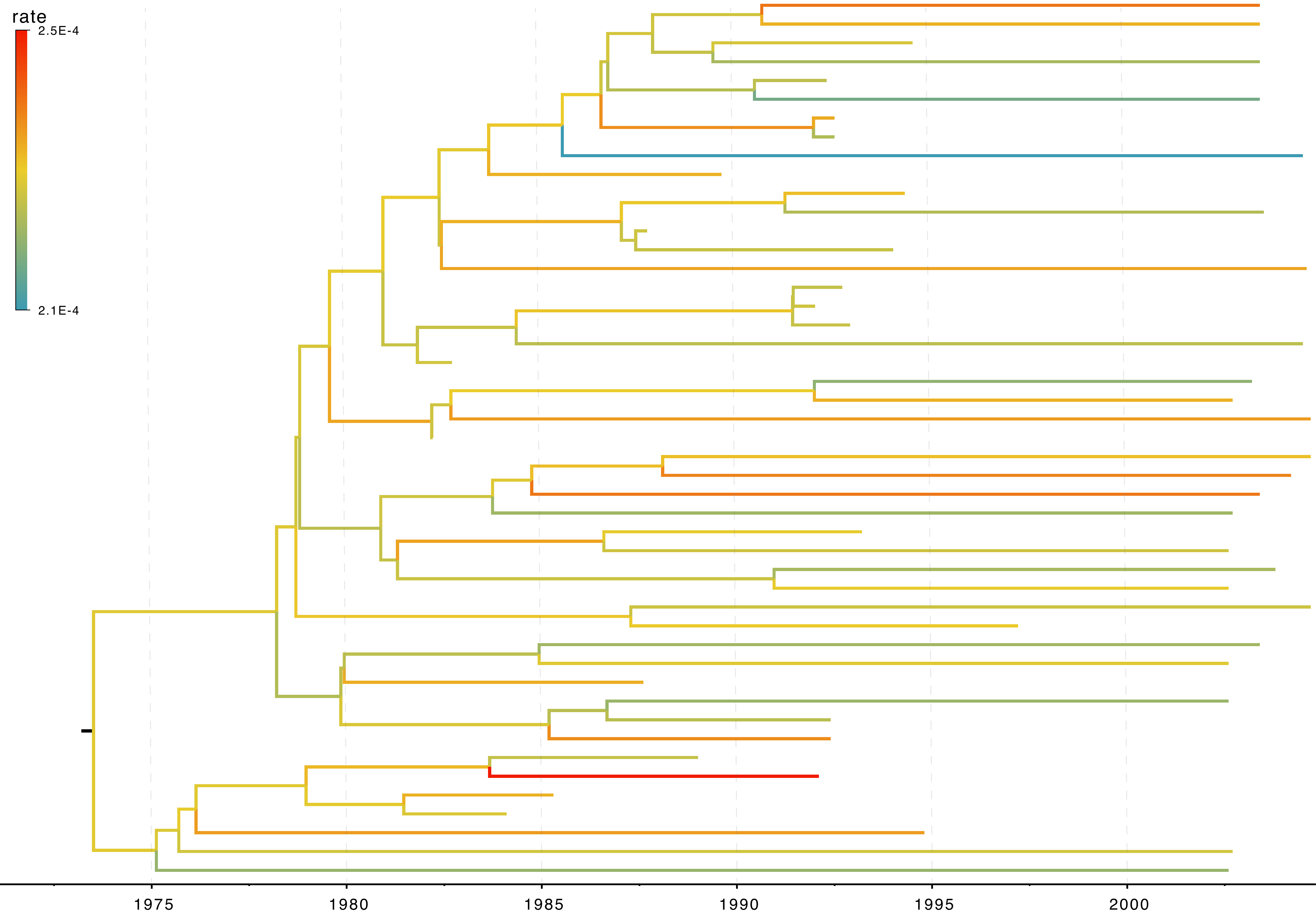}
		\caption{The RABV phylogeny explored in the example.
			Branches on the tree are color-coded by the posterior means of the branch-specific evolutionary rates.
		}
		\label{fig:RRV_tree}
	\end{center}
\end{figure}

\paragraph{Rabies virus}
Our analysis results in a posterior mean rate of
$2.12$ $(1.73, 2.51) \times 10^{-4}$ substitutions per site per year.
The estimated scale parameter has posterior mean $0.10$ $(0.00, 0.24)$.
Figure~\ref{fig:RRV_tree} shows the maximum clade credible evolutionary tree of the RABV example.
Our analysis estimates the date of the root of the tree to be
$1971.9$ $(1951.3, 1979.7)$.  This is slightly older than the estimate in \cite{biek2007high} and our $95\%$ Bayesian credible interval is wider.

\paragraph{Lassa virus}

\begin{figure}
	\begin{center}
		\includegraphics[width=1.0\textwidth]{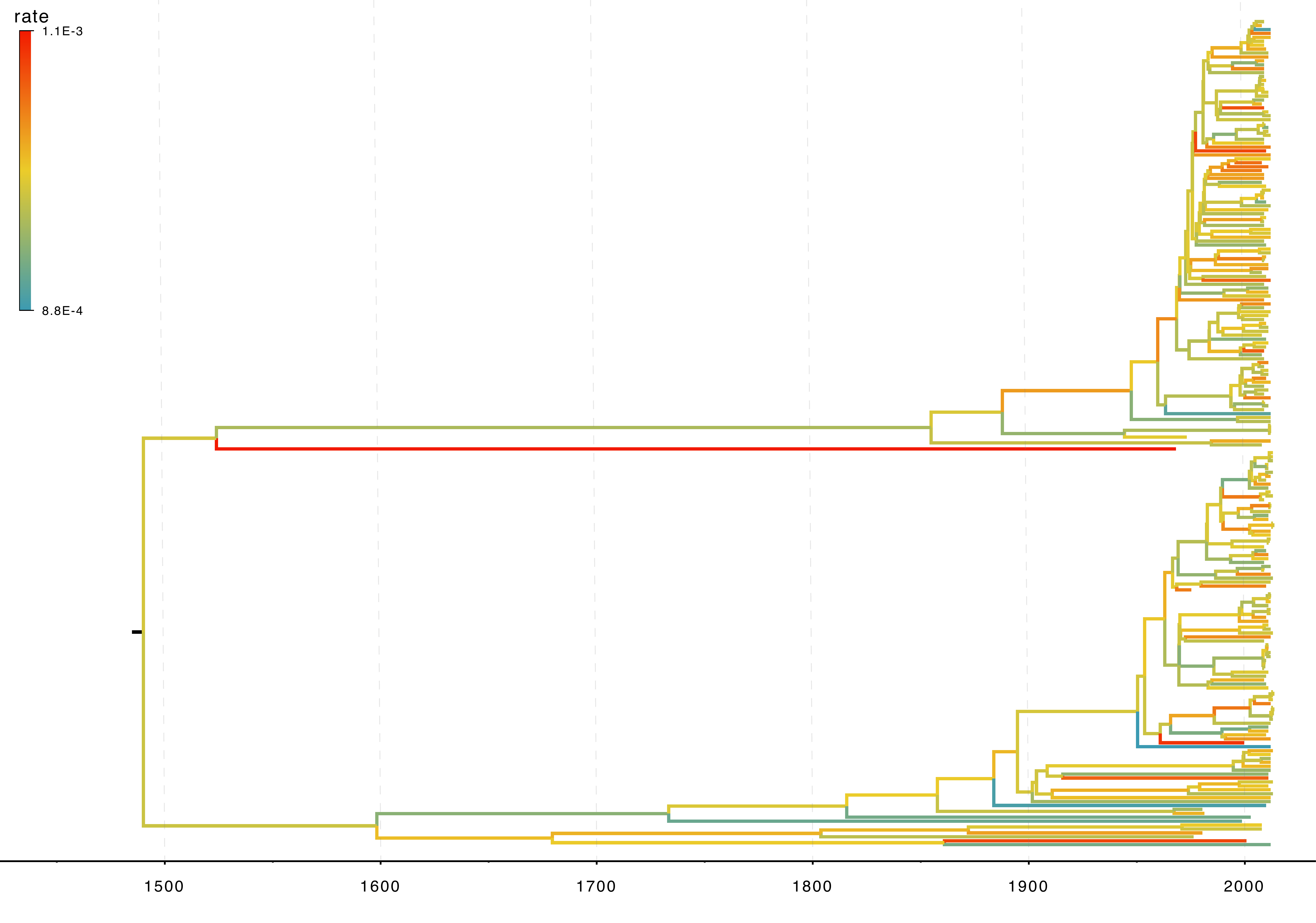}
		\caption{The LASV phylogeny explored in the example.
			Branches on the tree are color-coded by the posterior means of the branch-specific evolutionary rates.
		}
		\label{fig:Lassa_tree}
	\end{center}
\end{figure}

Our analysis yields a posterior mean rate of
$0.97$  $(0.81, 1.14)$ $\times 10^{-3}$ substitutions per site per year.
The estimated scale parameter has posterior mean $0.089$ $(0.035, 0.140)$.
Figure~\ref{fig:Lassa_tree} shows the maximum clade credible evolutionary tree of the LASV example.
The date of the root of the tree is inferred to be
$1434.0$ $(1059.0, 1601.7)$.  This agrees with
the finding by \cite{andersen2015clinical} that LASV is a long-standing human pathogen whose most recent common ancestor existed around six hundred years ago. 

\begin{figure}
	\begin{center}
		\includegraphics[width=1.0\textwidth]{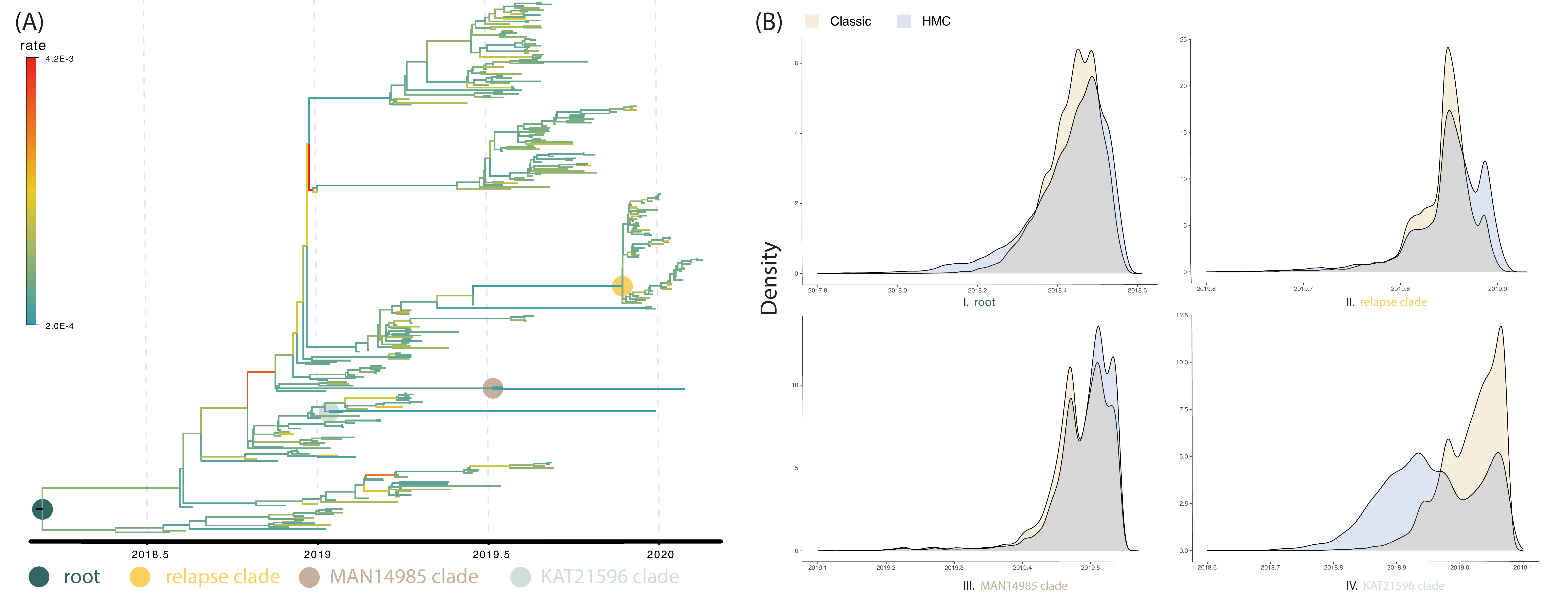}
		\caption{Kernel density estimation plot of the tMRCA distribution of four clades of interest on the EVD phylogeny.
			(A) The EVD phylogeny explored in the example.
			Branches on the tree are color-coded by the posterior means of the branch-specific evolutionary rates.
			We use four colored dots to indicate the four MRCA nodes of four clades of interest on the tree. 
			(B) The kernel density estimation plot of the tMRCA of the four nodes indicated in (A).
		}
		\label{fig:EVD}
	\end{center}
\end{figure}

\paragraph{Ebolavirus}
Our analysis yields a posterior mean rate $7.70$ $(6.63, 8.82) \times 10^{-4}$
substitutions per site per year.  The scale parameter has posterior mean $0.98$
$(0.64, 1.33)$.
Figure~\ref{fig:EVD} shows the maximum clade credible evolutionary tree of the EBOV example.
The inferred MCC tree shows a significant slow-down in evolutionary rate on the branch leading to the relapse clade that roughly spans over $5.3$ months, similar to the discovery from \cite{mbala2021ebola}.
However, our analysis reveals more variability in evolutionary rates compared to the original study.
The MAN4194 sequence that was collected from the individual with the
relapsed Ebola infection is basal to all other DRC sequences within the relapse clade.
Our analysis estimates the date of most recent common ancestor (MRCA) of the relapse clade (Figure~\ref{fig:EVD} B II) to be $2019.85$ $(2019.77, 2019.91)$.
This is similar to the estimate of \cite{mbala2021ebola}.
However, our analysis revealed a clearer bimodal posterior distribution that was previously missed.
Our estimated date of the MRCA of the MAN14985 clade (Figure~\ref{fig:EVD} B III) is $2019.49$ $(2019.42, 2019.54)$ and the estimated date of the
MRCA of the KAT21596 set (Figure~\ref{fig:EVD} B IV) is $2018.96$ $(2018.83, 2019.07)$.
%

\section{Discussion}
The confounding of evolutionary rate and time has imparted divergence time estimation with high uncertainty and low reliability of the inference. 
Nonetheless, much effort and improvement have shaped the molecular clock models to better characterize evolutionary rate heterogeneity along phylogenies \citep{thorne1998estimating, kishino2001performance, drummond2006relaxed, rannala2007inferring, lemey2010phylogeography, lartillot2016mixed, bletsa2019}.
%
%
We here introduce a linear-time transformation of the internal node height parameters into a ratio space with the aim to improve estimation efficiency under complex molecular clock models.
Naive transformation of the gradient of the log-likelihood from the original height space into the ratio space results in $\order{\nTips^2}$ computations.
To make the transformation scalable, we present linear-time algorithms that improve the performance of this transformation. 
With a slight modification, Algorithm~\ref{alg:post_order_gradient_log_determinant} builds upon Algorithm~\ref{alg:post_order_gradient_update} to calculate all derivatives of the log-determinant of the Jacobian matrix also in linear-time.
This collection of linear-time algorithms enables researchers to employ dynamic-based samplers (e.g., HMC) to sample the internal node heights and substantially improve inference.

When applying HMC on all dimensions in the ratio space, the sampler proposes a new set of values for the height parameter and all ratios which corresponds to a set of new values for all internal node heights in the original height space.
Alternatively, one may cycle HMC on subsets of dimensions from the ratio space in a Metropolis-within-Gibbs inference strategy such that in each iteration, HMC proposes new values to only a subset of dimensions.
For example, one possible choice of these subsets is to separately sample the root height and all the ratios (i.e.~one subset containing only the height parameter and one subset containing all ratio parameters).
Interestingly, each of the two subsets takes a full traversal for updating the gradient through Algorithms~\ref{alg:post_order_gradient_update} and \ref{alg:pre_order_gradient_update}
where the postorder traversal updates the gradient w.r.t.~all ratio parameters ($\nTips -2$ dimensional) and the preorder traversal updates only the height parameter (single dimensional). 
Therefore, sometimes it might be more computationally efficient to mix the classic univariable sampling kernels with HMC for the height dimension to benefit from the low computational load for learning the root height dimension.
For example, one may apply classic univariable samplers on the height dimension in ratio space instead of HMC.
In addition, one may apply classic univariable samplers on the original root height dimension such that with careful caching of the previous iteration, each proposal only needs updating two postorder partial likelihood vectors corresponding to the two immediate descendant branches from the root.
However, as illustrated by the WNV example, classic univariable samplers may suffer from the constraints on the node heights resulting in poor mixing in some dimensions (e.g.~several internal nodes close to the root in this case), where the mixture of samplers may lead to worse computational efficiency.
To investigate the univariable sampler's validity, we ran the chain 10x longer for the WNV example.
As expected, the trace plot of the longer chain exhibits a normal ``caterpillar'' shape that indicates both the validity and limitation of the univariable samplers.

The EBOV example employs a more general mixed-effects relaxed clock model with clade-specific fixed-effects and branch-specific random-effects.
The original study \citep{mbala2021ebola} incorporates rate variation into a strict molecular clock model by introducing a single parameter to capture fixed-effects from the clades of interest.
Their molecular clock model therefore has $2$ dimensions.
The mixed-effects model employed in this study now utilizes a $597$-dimensional parameter ($4$ dimensions for clade-specific fixed-effects with an intercept term, $592$-dimensions for branch-specific random-effects, and $1$ dimension for the scale parameter) to capture multiple sources of rate variation.
This more general mixed-effects model detects the same slow-down of the evolutionary rate of the branch leading to the relapse clade.
Interestingly, the relapse clade and the MAN14985 clade are monophyletic with
posterior probability approaching $1$ in our analyses whereas the KAT21596 clade is monophyletic with posterior probability $0.37$ compared to the posterior probability of $0.95$ in the original study.
The lower posterior probability estimate for the two sequences (KAT21596 and BTB4325) forming a monophyletic clade indicates a different mixture of tree topologies partly owing to the more general molecular clock model and potentially better mixing of node heights in each topology.
The difference in posterior probability of the KAT21596 clade further affects the multi-modal posterior distribution of tMRCA of the two sequences as in Figure~\ref{fig:EVD} (B).

Recent molecular clock models add additional dependence of evolutionary rate onto time \citep{aiewsakun2015time, ho2015time, membrebe2019bayesian} that bring in more biological insights into the time-dependency of the evolutionary rates in viral evolution.
However, such a dependence structure further complicates the confounding of evolutionary rate and time.
Fortunately, the complex dependence structure only affects the derivatives without influencing the ratio transformation or the HMC machinery and is therefore the reason Equation~\ref{eq:NodeHeightGradient} uses more general terms $\frac{\partial \bl{i}}{\partial \nodeT{i}}$, $\frac{\partial \bl{j}}{\partial \nodeT{i}}$, and $\frac{\partial \bl{k}}{\partial \nodeT{i}}$.

A caveat of the linear-time algorithms that are introduced
here is that they assume sampling dates are given and fixed.
Often, viral sequences are associated with various levels of uncertainty, not only in their associated metadata (e.g., sampling dates) but also with regard
to sequencing quality.
Typically, a quality control step removes unreliable sequences.
In a Bayesian framework, one may integrate out sampling date uncertainty through their support so that sampling dates become parameters of the model and are no longer fixed \citep{Pybus2012}.
The proposed algorithms and HMC machinery remain unaffected if one cycles between sampling all internal node heights and tip heights from their full conditional distributions.
However, the derivative w.r.t. the height parameter in the ratio space needs to consider contributions from the tip nodes when one samples all node heights (including variable tip heights) jointly.
Moreover, the anchor node and epoch constructions become variable and
need to be jointly updated with tip heights.
This remains an important avenue of future work.


\section{Materials and Methods}\label{sec:materials_and_methods}
We have implemented the algorithms in this manuscript within the development branch of the software package BEAST \citep{suchard2018} with likelihood computations off-loaded to the high-performance BEAGLE library \citep{ayres2019beagle}. 
We provide instructions and the BEAST XML files for reproducing these analyses on Github at \href{https://github.com/suchard-group/hmc_divergence_time_manuscript_supplement}{https://github.com/suchard-group/hmc\_divergence\_time\_manuscript\_supplement}.

\section{Acknowledgments}

The research leading to these results has received funding from the European Research Council under the European Union's Horizon 2020 research and innovation programme (grant agreement no.~725422 - ReservoirDOCS).
The Artic Network receives funding from the Wellcome Trust through project 206298/Z/17/Z.
MAS and XJ are partially supported by NIH grants U19 AI135995, R56 AI149004, R01 AI153044 and R01 AI162611.
GB acknowledges support from the Interne Fondsen KU Leuven / Internal Funds KU Leuven under grant agreement C14/18/094 and from the Research Foundation -- Flanders (`Fonds voor Wetenschappelijk Onderzoek -- Vlaanderen', G0E1420N and G098321N).
PL acknowledges support by the Research Foundation -- Flanders (`Fonds voor Wetenschappelijk Onderzoek -- Vlaanderen', G066215N, G0D5117N and G0B9317N).
JLT was supported by NSF DEB 1754142.

\newpage
\bibliographystyle{chicago}

\end{document}